# Comparison of photonic transversal RF spectral filters based on different optical microcombs

Mengxi Tan[1], Xingyuan Xu[1], Jiayang Wu[1], Roberto Morandotti[2], Arnan Mitchell[3], and David J. Moss[1]

[1] X. Xu, M. X. Tan, J. Wu, and D. J. Moss are with Centre for Micro-Photonics, Swinburne University of Technology, Hawthorn, VIC 3122, Australia. (Corresponding e-mail: dmoss@swin.edu.au).
[2] R. Morandotti is with INRS -Énergie, Matériaux et Télécommunications, 1650 Boulevard Lionel-Boulet, Varennes, Québec, J3X 1S2, Canada, and is a visiting professor with ITMO University, St. Petersburg, Russia, and also with the Institute of Fundamental and Frontier Sciences, University of Electronic Science and Technology of China, Chengdu 610054, China.
[3] A. Mitchell is with the School of Engineering, RMIT University, Melbourne, VIC 3001, Australia.

*Abstract*— **Integrated Kerr microcombs are emerging as a powerful tool as sources of multiple wavelength channels for photonic RF and microwave signal processing mainly in the context of transversal filters. They offer a compact device footprint, very high versatility, large numbers of wavelengths, and wide Nyquist bands. Here, we review recent progress on Kerr microcomb-based photonic RF and microwave reconfigurable filters, based both on transversal filter methods and on RF to optical bandwidth scaling. We compare and contrast results achieved with wide comb spacing combs (200GHz) with more finely spaced (49GHz) microcombs. The strong potential of optical micro-combs for RF photonics applications in terms of functions and integrability is also discussed.**

## I. INTRODUCTION

Nonlinear optics for all-optical signal processing has proven to be an extremely powerful approach, particularly when implemented in photonic integrated circuits based on highly nonlinear materials such as silicon [1, 2]. All optical signal processing functions include all-optical logic [3], demultiplexing from 160Gb/s [4] to over 1Tb/s [5], optical performance monitoring via slow light [6-7], all-optical regeneration [8,9], and others [10-16]. CMOS compatible platforms are centrosymmetric, and so nonlinear devices have been based on third order nonlinear processes including third harmonic generation [11,17-21] and, the Kerr nonlinearity ($n_2$) [1,2]. The efficiency of Kerr based all-optical devices depends on the waveguide nonlinear parameter, $\gamma = \omega\, n_2\, /\, c\, A_{eff}$ where $n_2$ is the Kerr nonlinearity and $A_{eff}$ is the waveguide effective area). Although silicon can achieve extremely high values of $\gamma$, it suffers from high nonlinear losses due to two-photon absorption (TPA) and the resulting free carriers [2]. Even if the free carriers are eliminated by p-i-n junctions, silicon's poor intrinsic nonlinear figure of merit (FOM = $n_2\,/\,(\beta\,\lambda)$, where $\beta$ is the TPA) of around 0.3 in the telecom band is far too low to achieve high performance. While TPA can be turned to advantage for some all-optical functions [22-24], for the most part silicon's low FOM in the telecom band is a limitation. This has motivated research on a range of alternate nonlinear platforms highlighted by perhaps by the chalcogenide glasses [25-35]. While offering advantages, these platforms are not compatible with the silicon computer chip industry – CMOS (complementary metal oxide semiconductor).

In 2008 new CMOS compatible platforms for nonlinear optics, including silicon nitride [35, 36] and Hydex [37-47] were introduced with negligible nonlinear absorption in the telecom band, a moderate nonlinear parameter and extremely high nonlinear figure of merit, which are ideal for micro-comb generation [47]. Following the first report of Kerr frequency comb sources in 2007 [48], the first integrated CMOS compatible integrated optical parametric oscillators were reported in 2010 [36, 37], and since then this field has exploded [47]. Many cutting-edge applications have been demonstrated based on CMOS-compatible micro-combs, ranging from filter-driven mode-locked lasers [49-52] to quantum physics [53-58]. Most recently, Kerr micro-combs have demonstrated their enormous potential for sources for ultrahigh bandwidth coherent optical fiber communications [59], optical frequency synthesis [60] and many other applications [61-68]. The success of these new CMOS platforms as well as other CMOS platforms such as amorphous silicon [69] and silicon rich silicon nitride (SRSN) [70] arises from a combination of their low linear loss, moderate to high nonlinearity and low or even negligible nonlinear loss (TPA).

Photonic RF techniques are an important application of all-optical signal processing. They have attracted great interest during the past two decades since they offer ultra-high RF bandwidths, low transmission loss and strong immunity to electromagnetic interference. They have found wide applications ranging from radar systems to communications [71-100]. There are a very wide variety of approaches to realizing photonic RF filters, including ones that map the optical filter response onto the RF domain [80-95], highlighted by on-chip (waveguide based) stimulated Brillouin scattering [83-85]. This approach has achieved extremely high performance in terms of RF resolution — as high as 32 MHz — and a stopband rejection >55 dB. Another key approach focuses on reconfigurable transfer functions for adaptive signal processing based on transversal filters [86-91]. These operate by generating weighted and progressively delayed replicas of the RF signal in the optical domain and then combining them upon photo-detection. Transverse filters are capable of achieving arbitrary RF transfer functions by simply changing the tap weights, and so are attractive for the implementation of advanced adaptive and dynamic RF filters. Typically, discrete laser arrays [92-94] or Bragg grating arrays [95,96] have been employed to supply the needed taps. However, while offering many advantages, these approaches result in significantly increased complexity as well as reduced performance due to the limited number of available taps. Alternative approaches, including those based on optical frequency comb sources achieved by electro-optic (EO) modulation, can help mitigate this problem, but they require cascaded high frequency EO [97-99] or Fabry-Perot EO [100] modulators that in turn require high-frequency RF sources. Of these approaches, EO comb-based methods have been in use the longest and have demonstrated many powerful functions.

Integrated optical Kerr frequency comb sources, or "microcombs" [36-52, 102-113] have come into focus as a fundamentally new and powerful tool due to their ability to provide highly coherent multiple wavelength channels with a very high degree of wavelength spacing control, from a single source. They originate via optical parametric oscillation in monolithic micro-ring resonators (MRRs), offer significant advantages over more traditional multi-wavelength sources for RF applications. They can greatly increase the capacity of communications systems and allow the processing of RF spectra for very wide range of advanced signal processing functions [113-132]. They have the advantage of being able to generate frequency combs with much higher spacings than electro-optic combs. In fact, EO combs and microcombs are in many ways complementary – the former focusing on finer wavelength spacings from 10's of MHz to 10 - 20 GHz, while the microcombs typically have much wider spacings from 10-20 GHz up to 100's of GHz and even terahertz spacing. Wider comb spacings yield much wider Nyquist zones for larger RF bandwidths, whereas finer spacings potentially can yield much larger numbers of wavelengths or RF "taps", at the expense of much more limited Nyquist zones and hence RF bandwidths. Micro-combs have the potential to provide a much higher number of wavelengths, an ultra-large FSR, as well as greatly reduced footprint and complexity. In particular, for RF transversal functions, the number of wavelengths dictates the available channel number of RF time delays, and thus with microcombs, the performance of RF filters and other systems such as beamforming devices can be greatly enhanced in terms of the quality factor and angular resolution, respectively. In addition, other approaches to filtering such as RF bandwidth scaling [126] have a given bandwidth for each wavelength channel, and so the total operation bandwidth (i.e., the maximum bandwidth of the input RF signal that can be processed) will depend on the number of wavelengths, and hence will be greatly enhanced with microcombs. Based on these advantages, a wide range of RF applications have been demonstrated, such as optical true time delays [118], transversal filters [122, 126], signal processors [115,123], and channelizers [125].

Here, we focus on RF and microwave spectral filters based on transversal filtering methods that exploit integrated Kerr microcomb devices. We review our work on fixed and tunable bandpass filters as well as less common filters such as gain equalizing, low pass etc.. We discuss the tradeoffs between using very widely spaced microcombs at 200GHz with recently reported [115, 118, 122, 126] record low FSR spaced microcombs at 49 GHz, based on soliton crystals. We review the recent advances of RF and microwave filtering functions made possible through the use of microcombs, highlighting their potential and future possibilities, contrasting the different methods and use of the differently spaced microcombs. The RF transversal filters based on 200GHz Kerr micro-combs achieved a high degree of versatility and dynamic reconfigurability. However, their relatively large comb spacing arising from the large free spectral range (FSR) of ~1.6 nm (~200 GHz) restricts the number of taps to less than 21 within the 30nm wide bandwidth of the C-band. This is an important consideration since this approach requires optical amplifiers and optical spectral shapers which are more readily available for wavelengths in the C-band. The limitation in the number of taps has limited the performance of micro-comb based transversal RF filters in terms of frequency selectivity, tuning resolution, dynamic versatility (filter shapes), as well as the performance of RF photonic signal processors. In order to improve on this, we demonstrated micro-comb-based photonic RF transversal filters with record high numbers of taps, featuring up to 80 wavelengths over the C-band [115,118,122,126]. This is the highest number of wavelengths so far reported for micro-comb-based RF transversal filters, enabled by a record low 49GHz-free-spectral-range integrated Kerr micro-comb source. This resulted in a $Q_{RF}$ factor for the RF bandpass filter of four times higher than the 200GHz FSR results [134]. Further, by programming and shaping the Kerr optical micro-comb, we achieve [122] RF filters with a high out-of-band rejection of up to 48.9 dB using Gaussian apodization, as well as a significantly improved tunable centre frequency covering the RF spectra range (from $0.05 \times FSR_{RF}$ to $0.40 \times FSR_{RF}$). We also demonstrate an adaptive photonic RF filter with highly reconfigurable 3-dB bandwidths (from 0.5 to 4.6 GHz) and arbitrary filter shapes. Our experimental results agree well with theory, verifying the feasibility of our approach towards the realization of high performance advanced adaptive RF transversal filters with potentially reduced cost, footprint, and complexity than other solutions.

## II. INTEGRATED KERR MICROCOMBS

The generation of microcombs is a complex process that generally relies on high nonlinear optical parameters, low linear and nonlinear loss as well as engineered anomalous dispersion. Diverse platforms have been developed for microcomb generation [47] such as silica, magnesium fluoride, silicon nitride, and doped silica glass. The MRRs used in the experiments reviewed here were fabricated using CMOS compatible fabrication processes, with Q factors of over 1.2 million and radii of ~592 μm and ~135 μm, corresponding to FSRs of ~0.4 nm (~49 GHz) and ~1.6 nm (~200 GHz), respectively [118, 127].

The MRRs used to generate the Kerr optical micro-combs are shown in Figure 1a for the 200GHz FSR combs and 1b for the 49GHz combs. They were both fabricated on a high-index doped silica glass platform using CMOS-compatible fabrication processes [37,38]. First, high-index (n = ~1.7 at 1550 nm) doped silica glass films were deposited using plasma enhanced chemical vapour deposition, then patterned by deep ultraviolet photolithography and etched via reactive ion etching to form waveguides with exceptionally low surface roughness. Finally, silica (n = ~1.44 at 1550 nm) was deposited as an upper cladding. The device architecture typically uses a vertical coupling scheme where the gap (approximately 200nm) can be controlled via film growth — a more accurate approach than lithographic techniques. The advantages of our platform for optical micro-comb generation include ultra-low linear loss (~0.06 dB·cm$^{-1}$), a moderate nonlinear parameter (~233 W$^{-1}$·km$^{-1}$), and in particular a negligible nonlinear loss up to extremely high intensities (~25 GW·cm$^{-2}$). Due to the ultra-low loss of our platform, the MRR features narrow resonance linewidths, corresponding to a quality factor of ~1.5 million (Fig. 1) for both MRRs. After packaging the device with fiber pigtails, the through-port insertion loss was as low as 0.5dB/facet, assisted by on-chip mode converters. The radius of the 49GHz (0.4nm) FSR MRR was ~592 μm, while the 200GHz FSR device had a radius of 135 μm. The 200GHz FSR device was the basis of our early work from the first report in 2008 on nonlinear optics in MRRs [39] followed by our work on micro-combs [37, 38]. In particular, the reports of microcombs in 2010 [36, 37] were the first reports of optical microcombs based on a fully integrated photonic chip platform, following the first report of microcombs in toroidal resonators [48]. Our work on true time delays [118] reported the first integrated microcomb able to generate soliton states, with an FSR < 100GHz. The record small optical free spectral range of the 49GHz FSR MRR greatly expanded the number of wavelengths (channels) available over the telecommunications band to as many as 80 wavelengths, or taps for microwave applications - much larger than previous reports.

To generate microcombs with the 200GHz device, the CW pump power is typically amplified and then the wavelength is swept from blue to red. When the detuning between the pump wavelength and MRR's cold resonance wavelength becomes small enough such that the intracavity power reaches a threshold, a modulation instability driven oscillation is initiated [47]. As the detuning is changed further, single-FSR spaced microcombs are generated. Achieving rigorous single soliton states in micro-combs via dissipative Kerr solitons (DKS), while well understood [103], is nonetheless very challenging. It requires both frequency and amplitude sweeping in both directions in order to "capture" the soliton states out of the initially chaotic states. Our initial work on microcomb applications were based on the 200GHz microcombs operating in a partially coherent state that, while not exhibiting rigorous soliton behaviour, was nonetheless a relatively low noise state that avoided the chaotic regime [47] and was more than adequate for microwave applications. The spectrum shown in Figure 1 is typical of the 200GHz spectra that we employed for our initial microwave work, and clearly shows that it is not a rigorous soliton state. Our work on the 49GHz FSR combs is based on soliton crystals [104, 118, 122, 126] resulting from mode crossings in the C-band. Soliton crystals are in fact much easier and robust to generate than DKS states and indeed even our partially coherent 200GHz states. They are achievable through simple manual tuning of the pump laser wavelength which is impossible with any other state. One reason for this is that the internal cavity optical energy of the soliton crystal state is very close to that of the chaotic state, and so when transitioning from one to the other, there is a very small shift in internal energy and hence resonant wavelength jump. It is this "jump" in both internal energy and resonant wavelengths of the microresonator that primarily is responsible for the challenge in realizing DKS states. The efficiency of the soliton crystal states – ie., the energy in the comb lines relative to the pump wavelength – is also much higher than the DKS states (the single soliton states in particular). The tradeoff is that the output spectrum of soliton crystals is not flat but exhibits a characteristic "curtain-like" pattern. While this typically requires a spectral flattening step, it has not posed a fundamental barrier to achieving a wide range of high performance RF and microwave processing functions. Soliton crystals also result from mode-crossings in the resonator, which can be challenging to design, rather than simply requiring the anomalous dispersion that DKS states need.

To generate soliton crystal Kerr micro-combs, we adjusted the polarization and wavelength of the pump light to one of the TE resonances of the MRR typically the resonance at ~1553.2 nm [118, 122] with the pump power set at ~30.5 dBm. When the detuning between the pump wavelength and the cold resonance became small enough, such that the intra-cavity power reached a threshold value, modulation instability (MI) driven oscillation was initiated. Primary combs were thus generated with the spacing determined by the MI gain peak — mainly a function of the intra-cavity power and dispersion. As the detuning was changed further, distinctive 'fingerprint' optical spectra were observed (Fig. 2). These spectra are similar to what has been reported from spectral interference between tightly packed solitons in the cavity — so called "soliton crystals" [37, 38]. An abrupt step in the measured intracavity power (Fig. 3(b)) was observed at the point where these spectra appeared, as well as a dramatic reduction in the RF intensity noise (Fig. 3(c)). Together with the spectra shapes, we take these observations as indicative of soliton crystal formation [38], although to conclusively demonstrate this one would need to perform time resolved pulse autocorrelation measurements. The key issue for our experiments was not the specific nature of the micro-comb state of oscillation so much as the low RF noise and high coherence, which were relatively straightforward to achieve through adiabatic pump wavelength sweeping. We found that it was not necessary to achieve any specific state, including either soliton crystals or single soliton states in order to obtain high performance — only that the chaotic regime [47] should be avoided. Nonetheless the soliton crystal states provided the lowest noise states of all of our microcombs, and was used as the basis for a microwave oscillator with low phase-noise [116]. This is important since there are a much wider range of coherent low RF noise states that are more readily accessible than any specific soliton related state [47].

## III. RF TRANSVERSAL FILTER THEORY

The transversal structure, similar to the concept of finite impulse response digital filters, is a fundamental tool for photonic RF signal processing. With the proper design of the tap weights, any transfer function can be arbitrarily realized for different signal processing functions such as bandpass filters, differentiators and Hilbert transformers [127]. The transfer function of the microwave photonic filter (MPF) can be described as (Figure 3):

$$H(\omega) = \sum_{n=0}^{N-1} a_n e^{-j\omega nT} \tag{1}$$

where $\omega$ is the angular frequency of the input RF signal, $N$ is the number of taps, $a_n$ is the tap coefficient of the $n_{th}$ tap, and $T$ is the time delay between adjacent channels (see Fig. 3). The Nyquist frequency of the MPF is given by $f_{Nyquist} = 1/2T$. By imposing calculated tap coefficients $a_n$ on each wavelength channel, reconfigurable MPFs with arbitrary spectral transfer functions can be implemented. Figure 3 shows a schematic diagram of the MPFs based on wavelength division multiplexing. The large number of wavelengths offered by the microcombs provides significant advantages in the performance of transversal signal processors. The progressively delayed RF replicas are termed "taps", and the number of taps directly determines the resolution, or the $Q_{RF}$ factor, of the system. By properly setting the tap coefficients, a reconfigurable transversal filter with diverse transfer functions can be achieved, including Hilbert transformers [115, 124], differentiators [123], and bandpass filters [120, 122, 127].

As Fig. 3 shows, the core of the microcomb-based transversal structure mainly contains two operations: broadcast and then delay. The input RF signal is first broadcast onto equally spaced wavelength channels via optoelectronic modulation to generate multiple replicas of the RF input. The replicas are then transmitted through a dispersive medium to acquire wavelength-sensitive delays. Second-order dispersion, which yields a linear relationship between the wavelength and the delay, is generally employed to progressively separate the replicas. The delayed signals can be either individually converted into electronic signals or summed upon photodetection. The former can be implemented through wavelength demultiplexers and photodetector arrays to perform the function of RF true time delays [118], while the latter can serve as transversal filters for RF signal processing [127].

Figure 4 shows the experimental setup for a system based on an integrated Kerr optical comb source with a 200GHz FSR (Figure 4a) and 49GHz FSR (Figure 4b). The main difference between the two experimental setups is the use of a second waveshaper for the 49GHz FSR comb, used to pre-flatten the scallop-shaped comb spectrum that is a distinctive hallmark of soliton crystal combs. The micro-comb, in both cases generated by the on-chip MRR, is then amplified and fed to either a single (200GHz comb) or dual (49GHz comb) waveshapers for channel equalization and weighting. The power of each comb line is manipulated by the waveshapers to achieve appropriately weighted tap coefficients. To increase the accuracy, we adopted a real-time feedback control path to read and shape the power of the comb lines accurately. The comb line powers were first detected by an optical spectrum analyser and then compared with the ideal tap weights. This allowed us to generate an error signal that was fed back into the waveshaper to calibrate the system and achieve accurate comb shaping. The processed comb lines were then divided into two parts according to the algebraic sign of the tap coefficients, and then fed into a 2×2 balanced MZM biased at quadrature. The 2×2 balanced MZM could simultaneously modulate the input RF signal on both positive and negative slopes, thus yielding replicas of the input RF signal with phase and tap coefficients having either algebraic sign.

For the 200GHz comb based system, the modulated signal produced by the MZM then went through ~2.122-km of standard SMF, where the dispersion was ~17.4 ps/(nm·km), corresponding to a minimum time delay $T$ of ~59 ps between adjacent taps (with the channel spacing of the time delay lines set equal to the FSR of the MRR), yielding a Nyquist frequency of ~8.45 GHz for the MPF. We note that the operational bandwidth of the MPF was determined by the Nyquist frequency, which could be easily enlarged by decreasing the time delay and, owing to the large FSR of the compact MRR, could potentially reach over ~100 GHz. Finally, the weighted and delayed taps were combined upon detection and converted back into RF signals at the output.

For the 49GHz comb based filter, the signal went through ~5-km of standard single mode fibre (SMF) to provide the progressive tap delays. The dispersion of the fibre was the same (17ps/nm/km), yielding a time delay $T$ of ~34.8 ps between adjacent taps, yielding an operation bandwidth (i.e., the Nyquist frequency, half of $FSR_{RF}$) of ~14.36 GHz for the transversal filter. This operation bandwidth can be easily enlarged by decreasing the time delay (e.g., using a shorter spool of SMF), at the expense of a reduced tuning resolution. However, the maximum operational bandwidth of the transversal filter is limited by the comb spacing. Significant crosstalk between adjacent wavelength channels (or taps) occurs for RF operation beyond 24.5 GHz — half of the microcomb's spacing 49 GHz. This issue can be addressed by employing a microcomb source with a larger comb spacing, although at the expense of providing fewer comb lines/taps across the C-band. Finally, as for the 200GHz based system, the weighted and delayed taps were combined and converted back into RF domain via a high-speed photodetector (Finisar, 40 GHz bandwidth).

## IV. FIXED FILTERS

Our initial work on micro-comb based MPFs was based on a 200GHz microcomb. We achieved noticeably better performance of some parameters such as the quality factor ($Q_{RF}$), of the bandpass filter (the ratio of its 3-dB bandwidth to $T$) as a result of the large number of taps provided by the optical comb source. In order to establish a benchmark for the bandpass filter, we used uniform tap coefficients. Figure 5 shows a simulated all-ones MPF with different numbers taps for the 200GHz FSR comb, and as can be seen, the 3-dB bandwidth decreases greatly as the tap number increases. The shaped optical combs exhibit a good match between experiment (red solid line) and theory (green crossing). The RF response of the all-ones MPF was characterized by a vector network analyser (VNA, Anritsu 37369A) (Fig. 5) confirming the increase in $Q_{RF}$ factor when expanding the tap number from 4 to 20.

By comparison, the 49GHz FSR combs yielded far superior performance in frequency selectivity, or resolution of the transversal filters, brought about by the large number of taps and represented by the $Q_{RF}$ factor. Figure 6 shows the micro-comb wavelengths used to implement low-pass sinc filters featuring equal tap weights (i.e., $h_{sinc}(n)=1$) for different numbers of taps. We measured the 3-dB bandwidth ($BW_{sinc}$) to calculate the corresponding $Q_{RF}$ factor ($Q_{RF} = BW_{sinc} / FSR_{RF}$). In the experiments, we selected different numbers of taps ranging from 2 to 80. The corresponding optical spectra are shown in Fig. 6. The RF transmission spectra of the sinc filter (Fig. 6(a)) were measured by a vector network analyser (VNA, Anritsu 37369A), and showed good agreement with theory (Fig. 6(b)). The measured $BW_{sinc}$ decreased from 3.962 to 0.236 GHz when the tap number was increased from 2 to 80, indicating a greatly enhanced $Q_{RF}$ of up to 73.7 with 80 taps — four times larger than the 200GHz results.

Figure 7 shows the dependence of the RF filter Q factor and bandwidth on the number of taps, contrasting the results obtained with the 200GHz device versus the 49GHz device. In advancing from 20 taps for the 200GHz device to 80 taps for the 49GHz device, the Q factor is improved from around 20 to almost 80, with a corresponding reduction in bandwidth from about 900 MHz to under 200MHz. Perhaps just as important is the out-of-band crosstalk. For the 200GHz device this is comparatively quite high at -10dB – probably too high for most practical applications. However, for the 49GHz device this drops to below -25dB which is a very significant improvement, rendering this device appropriate for practical applications. The theoretical $BW_{sinc}$ and $Q_{RF}$ as a function of the tap number are shown in Fig 7 as well as the experimental $Q_{RF}$ factor and 3-dB bandwidth for the 200GHz and 49GHz devices. As can be seen, $Q_{RF}$ increases linearly with the tap number and the 3-dB bandwidth decreases greatly as the tap number increases, further confirming the significant improvement of the frequency selectivity or resolution (reflected by $Q_{RF}$) brought about by the large tap number used here. The shaped optical combs (Fig. 6) also exhibit a good match between experiment (red solid line) and theory (green crossing), indicating that the comb lines were accurately shaped.

Finally, we note that it is not always a case of a smaller FSR comb being better. The 200 GHz devices are capable of reaching RF bandwidths, particularly in tunability, approaching 100GHz, or the Nyquist zone, whereas the 49GHz devices are limited to about 25GHz. We discuss tunable RF filters in the next section.

## V. TUNABLE FILTERS

Tunable bandpass filters are an important application in the RF signal processing field. While in conventional schemes this is generally obtained by physically varying the physical time delay, here we achieve this by only varying the tap coefficients. This illustrates the versatility and reach of micro-comb transversal filters, greatly reducing the complexity and instability of the system. We begin by focusing on the 200GHz FSR based filters, although the approach is the same for the 49GHz device. We employ the Remez algorithm [132] to determine the tap coefficients for the bandpass MPF with different center frequencies. Theoretically, the center frequency ($f_c$) and transition bandwidth ($BW_{tr}$) are critical to determining the $Q_{RF}$ factor and the required taps, and so we divide the design of the tunable MPF into two steps. First, the required dynamic range of the generated comb (i. e., the difference between the maximum and minimum power of the comb lines) and the necessary number of taps are calculated as functions of $f_c$ and $BW_{tr}$. We account for practical limitations such as a dynamic range of < 20 dB and a tap number < 20 (for the 200GHz based combs) in order to determine the available sets of $f_c$ and $BW_{tr}$. Secondly, we calculated the bandwidth and $Q_{RF}$ factor such that the optimized sets of $f_c$ and $BW_{tr}$ were subject to these practical restrictions. As a result, we are able to achieve a tunable MPF with optimized performance by just re-programming the tap weights, without the need for physical tunable delay lines. Thus, the advantages enabled by the use of Kerr combs are clear. In turn, this allowed a large number of wavelengths, providing a large number of taps that resulted in improved $Q_{RF}$ factors of the MPFs. For the 200GHz based device, we select 5 sets of tap coefficients resulting in 5 different center frequencies for the tunable MPF. Figure 8 verifies theoretically that the designed tunable MPFs could achieve acceptable performance. Figure 8 shows that the filter is able to achieve a tuning range from 2GHz to 6GHz with a bandwidth of about 1GHz. Figure 8 also shows the corresponding results for the 49GHz device, showing a dynamic tuning range from 2GHz to about 14GHz and with a bandwidth of 533MHz.

## VI. VERSATILE AND HIGH PERFORMANCE FILTERS

The large number of wavelengths provided by the 49GHz comb opens up the possibility of realizing both very high performance filters, in terms of resolution and out-of-band crosstalk, as well as versatility for providing almost arbitrary filter shapes. Figure 9 shows the filter weights and resulting filter spectral profile of a gaussian appodized bandpass filter using 80 taps. To improve the performance of the transversal filter in terms of out-of-band rejection, a Gaussian apodization was applied to the sinc filter [122], which was achieved by modifying the original discrete impulse response (i.e., tap weights) to the product of itself and a Gaussian function, given by

$$h_{Gau}(n) = h_{sinc}(n) \cdot e^{-\frac{(n-40.5)^2}{2\sigma^2}} \qquad (1)$$

where $\sigma$ is the root mean square width of the Gaussian function. Figure 9 shows the shaped comb spectra and the corresponding RF transmission spectra. As a function of decreasing $\sigma$, the main-to-secondary sidelobe ratio (MSSR) of the sinc filter increased from 26.4 to 48.9 dB, although at the expense of a deteriorated $Q_{RF}$ factor. While illustrating the capability of our transversal filter to achieve high out-of-band rejection, it also indicates that a large number of taps would be needed if both a very high $Q_{RF}$ factor and MSSR are simultaneously required. With the ability to offer a large number of taps — potentially over 200 in the C+L bands, micro-combs serve as a highly attractive approach to meet these demands. We note that the experimental results for $\sigma = 10$ were limited by our measurement noise floor of the VNA, and that up to 74.5 dB can in principle be achieved. Finally, Figure 10 shows RF filters for variable transmission across the passband (for gain equalizing functions etc) as well as a variable bandpass filter with the passband varying from 500MHz to over 4.6GHz. All of these filtering functions were achieved solely by varying the wavelength tap weights, with no changes in any of the physical parameters otherwise.

The results reported here highlight the advantages of using Kerr micro-combs as the basis for RF photonic microwave transversal filters. It also illustrates the tradeoffs between using widely spaced microcombs (200GHz) and record low spaced microcombs (49GHz) in terms of performance results. The greater number of lines supplied by the 49GHz comb (80 versus 20 for the 200GHz device) yield significantly better filter performance in terms of resolution, or Q-factor, tunability and cross-talk. On the other hand the 49GHz device is more limited in operational bandwidth, being restricted to roughly the Nyquist zone of 25GHz. The 200GHz device is able to reach RF frequencies that are well beyond what conventional electronic microwave technology can achieve.

## VII. CONCLUSION

Microcombs have shown their powerful capabilities for RF signal processing due to their large number of comb lines and compact footprint. We believe that they will bring further benefits to RF photonics, particularly in two aspects. First, the coherent nature of the soliton states will enable more advanced RF functions such as wideband frequency conversion and clock generation. Secondly, globally established CMOS fabrication platforms, which can perform hybrid integration of the microcomb source and III-V devices, will potentially enable monolithic integration of the entire RF system.

We demonstrate record performance and versatility for microcomb-based photonic RF transversal filters by employing 200GHz and 49GHz-FSR integrated optical micro-comb sources capable of providing record high numbers of wavelengths, or taps — up to 80 over the C-band- as well as a very high comb wavelength spacings for ultrahigh RF frequency applications. We achieve $Q_{RF}$ factors that are four times larger for the 49GHz devices compared with filters based on the 200GHz combs, as well as a high out-of-band rejection of up to 48.9 dB using Gaussian apodization. We also report RF filters with tunable centre frequencies achieved by only varying the tap weights. We cover the RF spectral range (from $0.05 \times FSR_{RF}$ to $0.40 \times FSR_{RF}$, or 1.4 to 11.5 GHz), as well as tunable 3dB bandwidths ranging from 0.5 to 4.6 GHz. Finally, we demonstrate highly reconfigurable filter shapes with positive and negative slopes across the passband as well as rectangular bandpass filter shapes. The experimental results agree well with theory, verifying that our transversal filter is a competitive approach towards achieving advanced adaptive RF transversal filters with broad operational bandwidths, high frequency selectivity, high reconfigurability, and potentially reduced cost and footprint, all of which are critical issues for modern radar and communications systems.

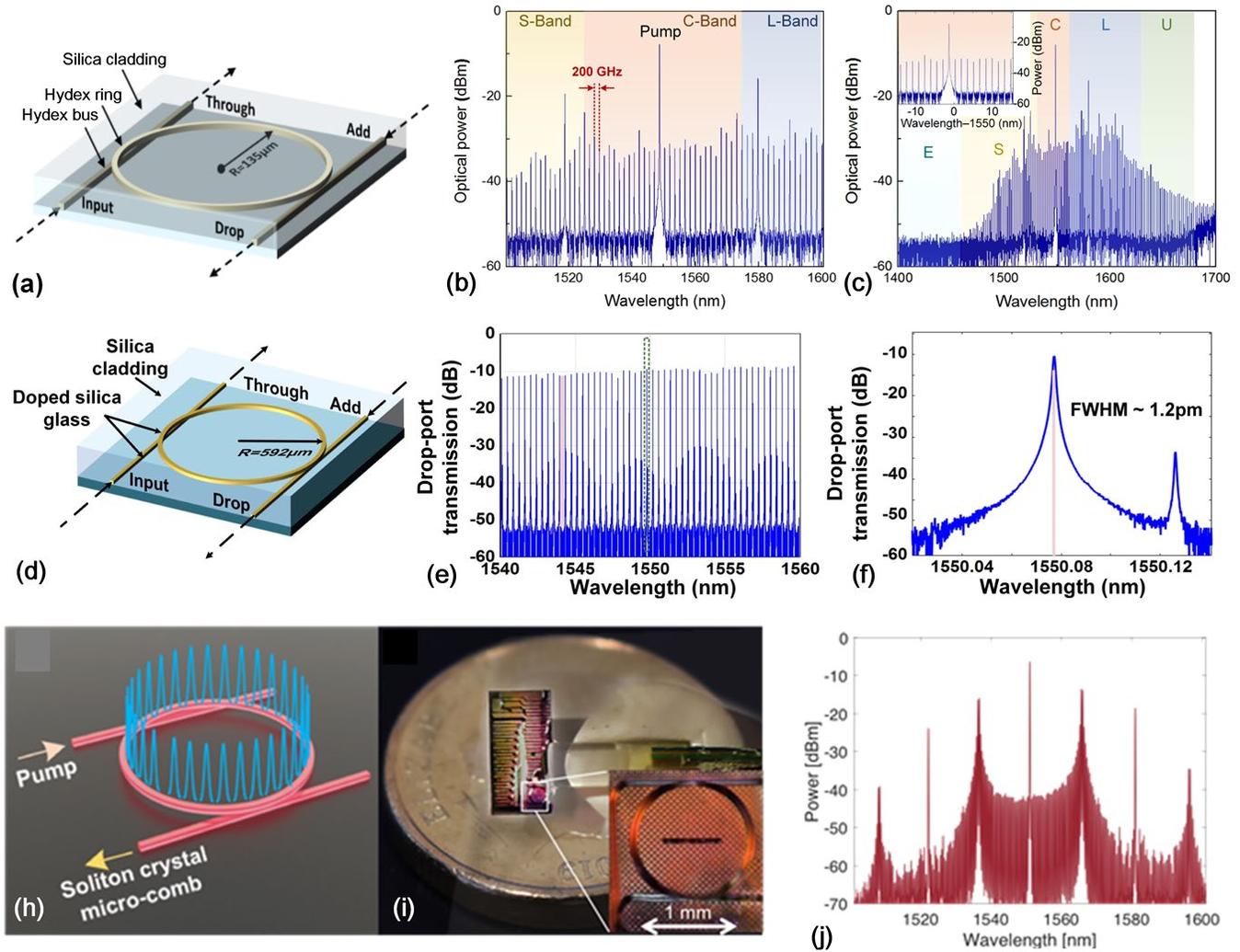

Fig. 1. Schematic illustration of the integrated MRRs for generating the Kerr micro-comb for both (a-c) 200GHz FSR combs and (d-j) 49GHz combs. (b, c)Optical spectra of the micro-combs generated by 200GHz MRR with a span of (b) 100 nm and (c) 300 nm. (j) Optical spectra of the micro-combs generated by 50GHz MRR with a span of 100 nm. (e) Drop-port transmission spectrum of the integrated MRR with a span of 5 nm, showing an optical free spectral range of 49 GHz. (f) A resonance at 193.294 THz with full width at half maximum (FWHM) of 124.94 MHz, corresponding to a quality factor of 1.549×106.

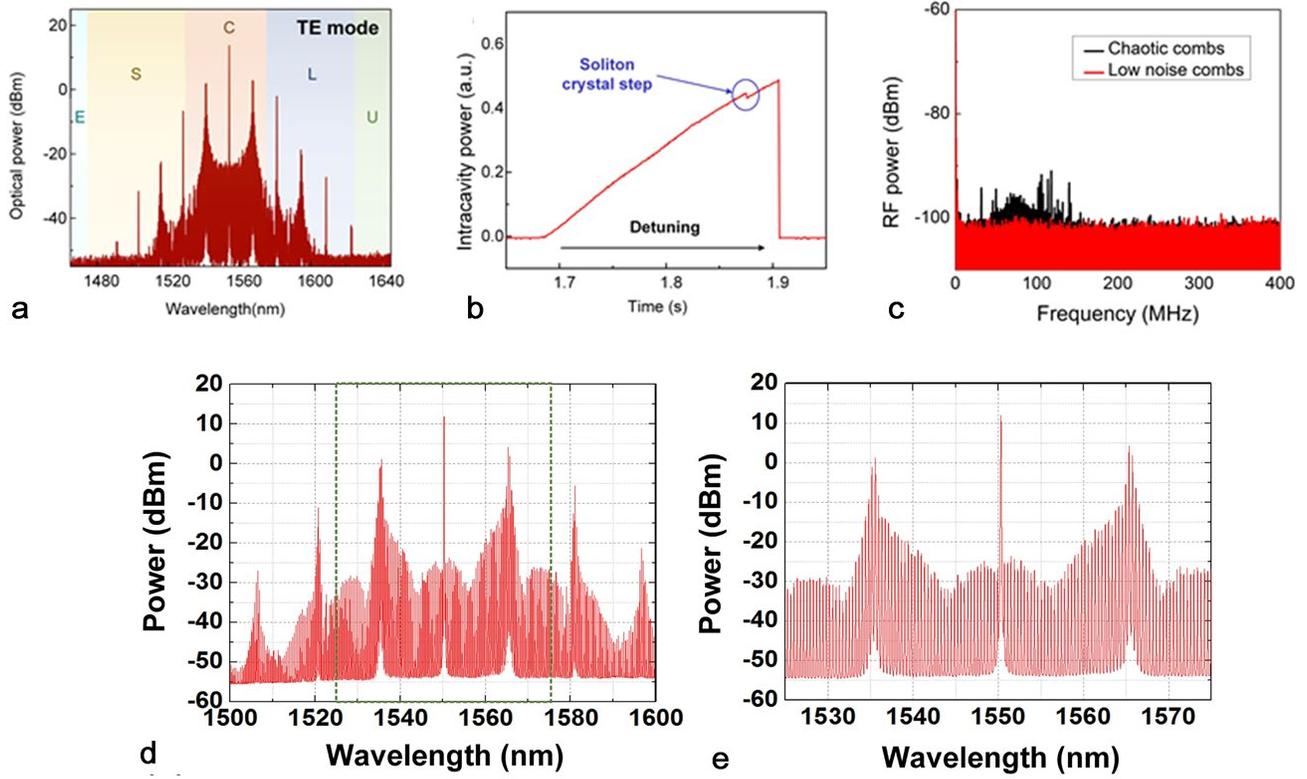

Fig. 2. (a, d, e) Optical spectra of various soliton crystal microcombs. (b) Optical power output versus pump tuning, showing the very small power jump at the onset of soliton crystal combs. (c) Transition from high RF noise chaotic state to low noise state of the soliton crystal comb.

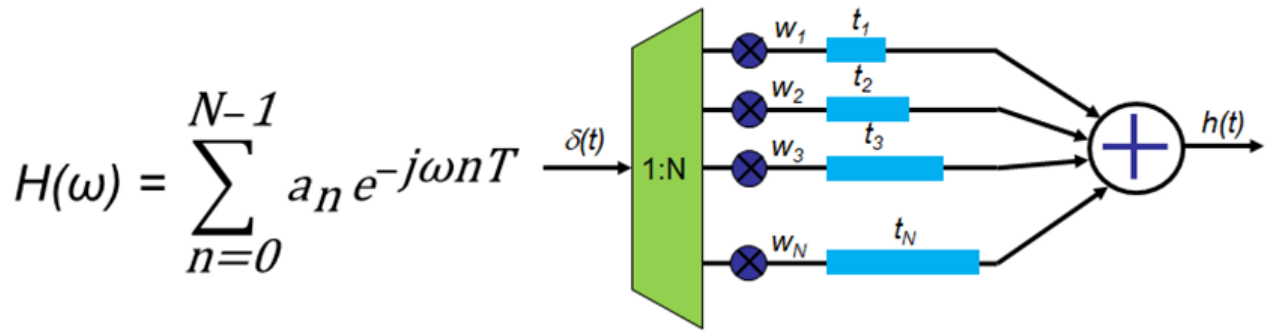

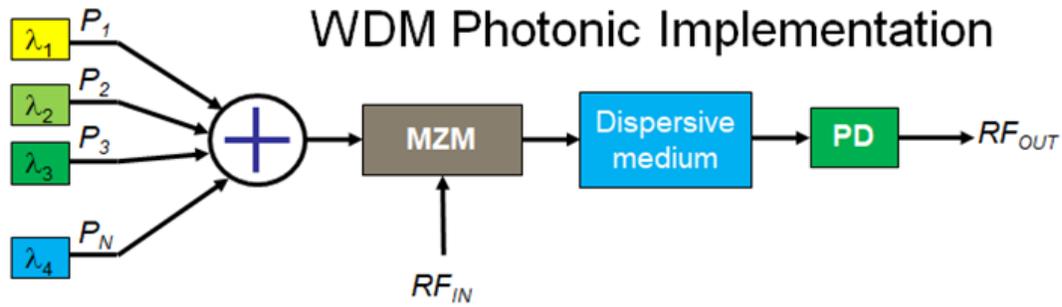

Fig. 3. Theoretical schematic of the principle of transversal filters using wavelength multiplexing. MZM: Mach-Zehnder modulators. PD: photo-detector.

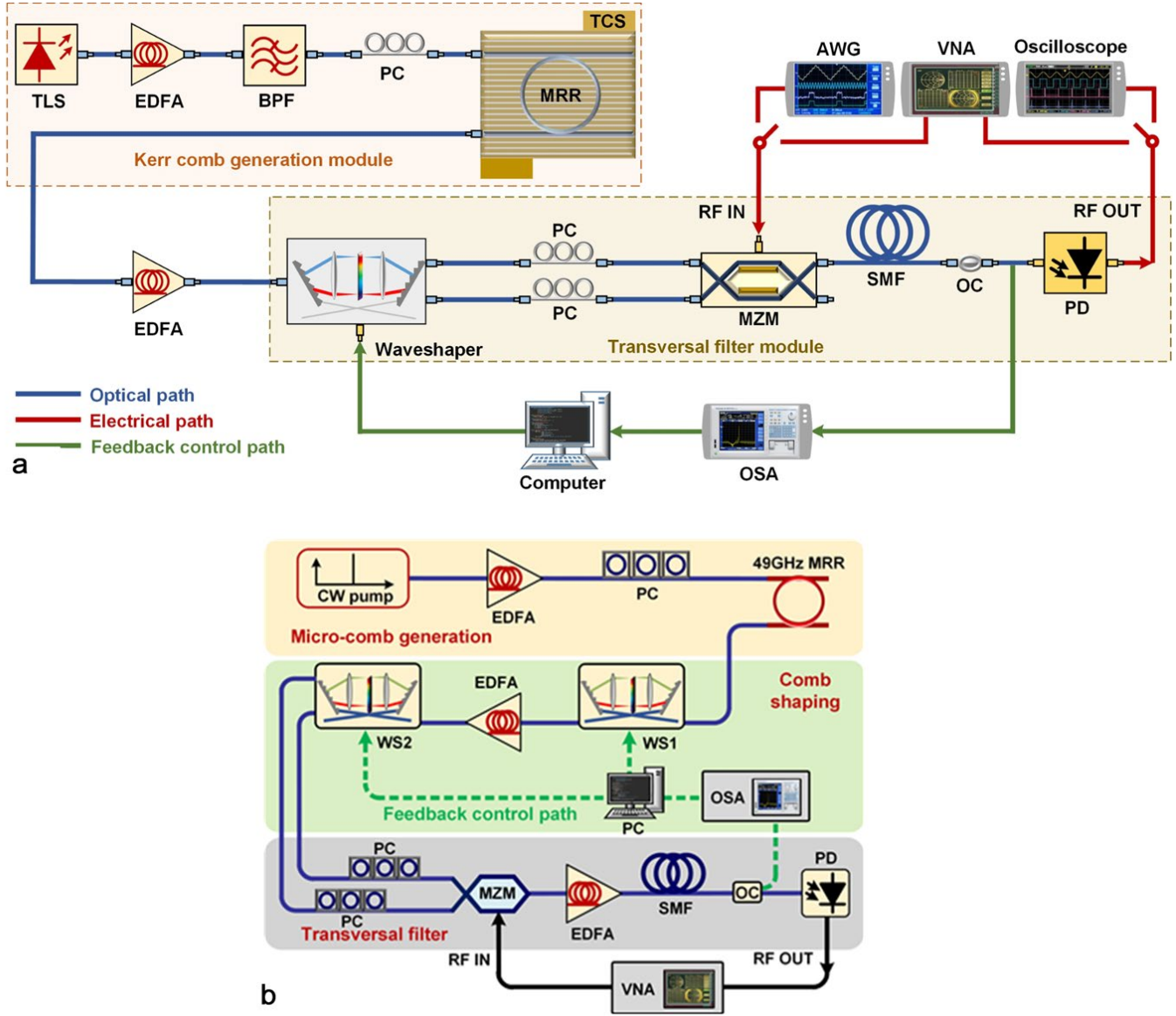

Fig. 4. Experimental schematic for RF transversal filters based on 200GHz microcomb (top) and 49GHz microcomb (bottom). TLS: tunable laser source. EDFA: erbium-doped fiber amplifier. PC: polarization controller. BPF: optical bandpass filter. TCS: temperature control stage. MRR: micro-ring resonator. WS: WaveShaper. OC: optical coupler. SMF: single mode fibre. OSA: optical spectrum analyzer. AWG: arbitrary waveform generator. VNA: vector network analyser.

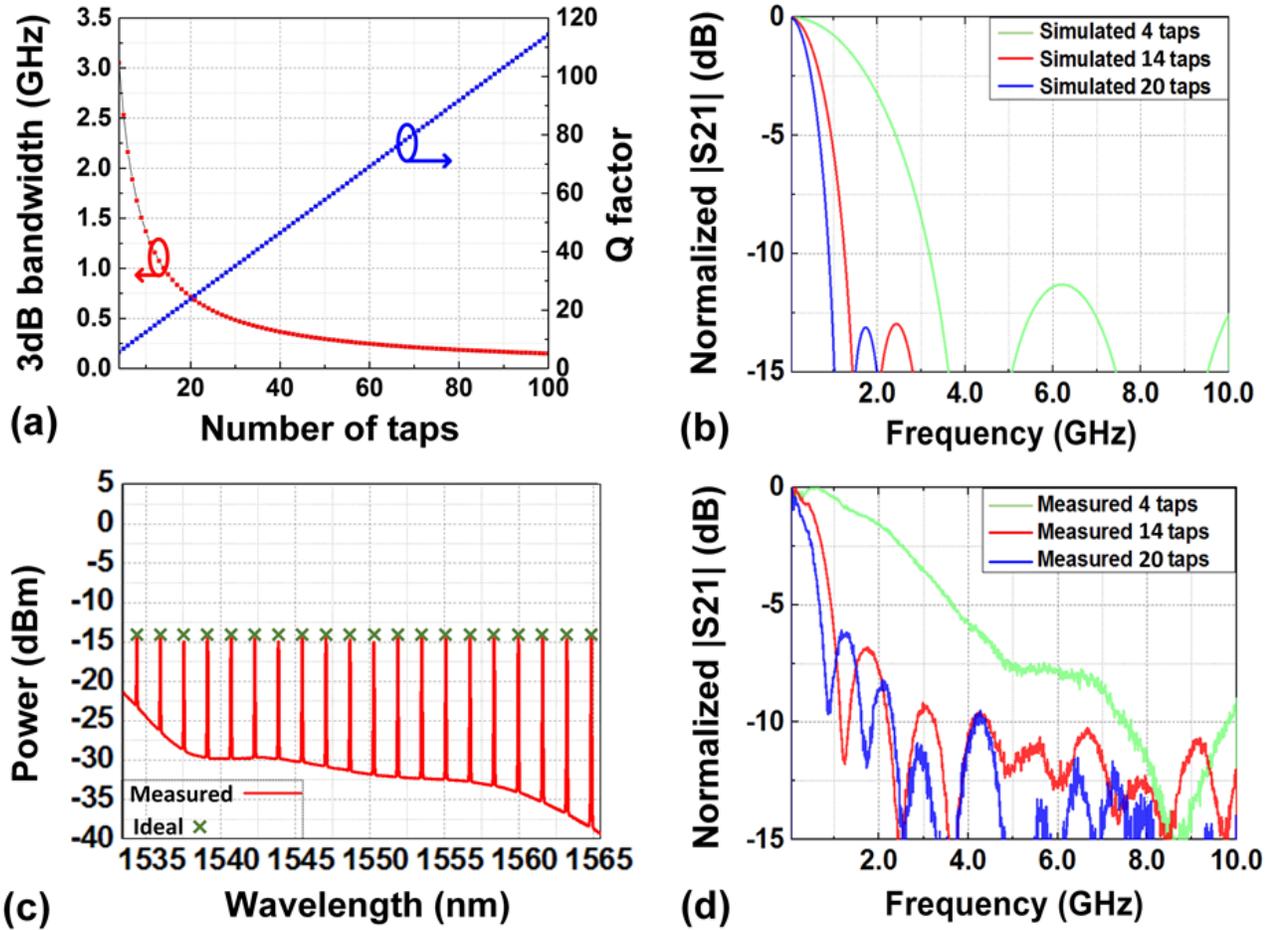

Fig. 5. Fixed RF lowpass filters with uniform tap weights. (a) Correlations between the number of taps and the Q factor, 3dB bandwidth of the all-ones MPF. (b) Simulated RF transmission spectra of an all-ones MPF with different number of taps. (c) Measured optical spectra (red solid) of the shaped optical combs and ideal tap weights (green crossing) for the all-ones MPF. (d) Measured RF transmission spectra of the all-ones MPF with different number of taps.

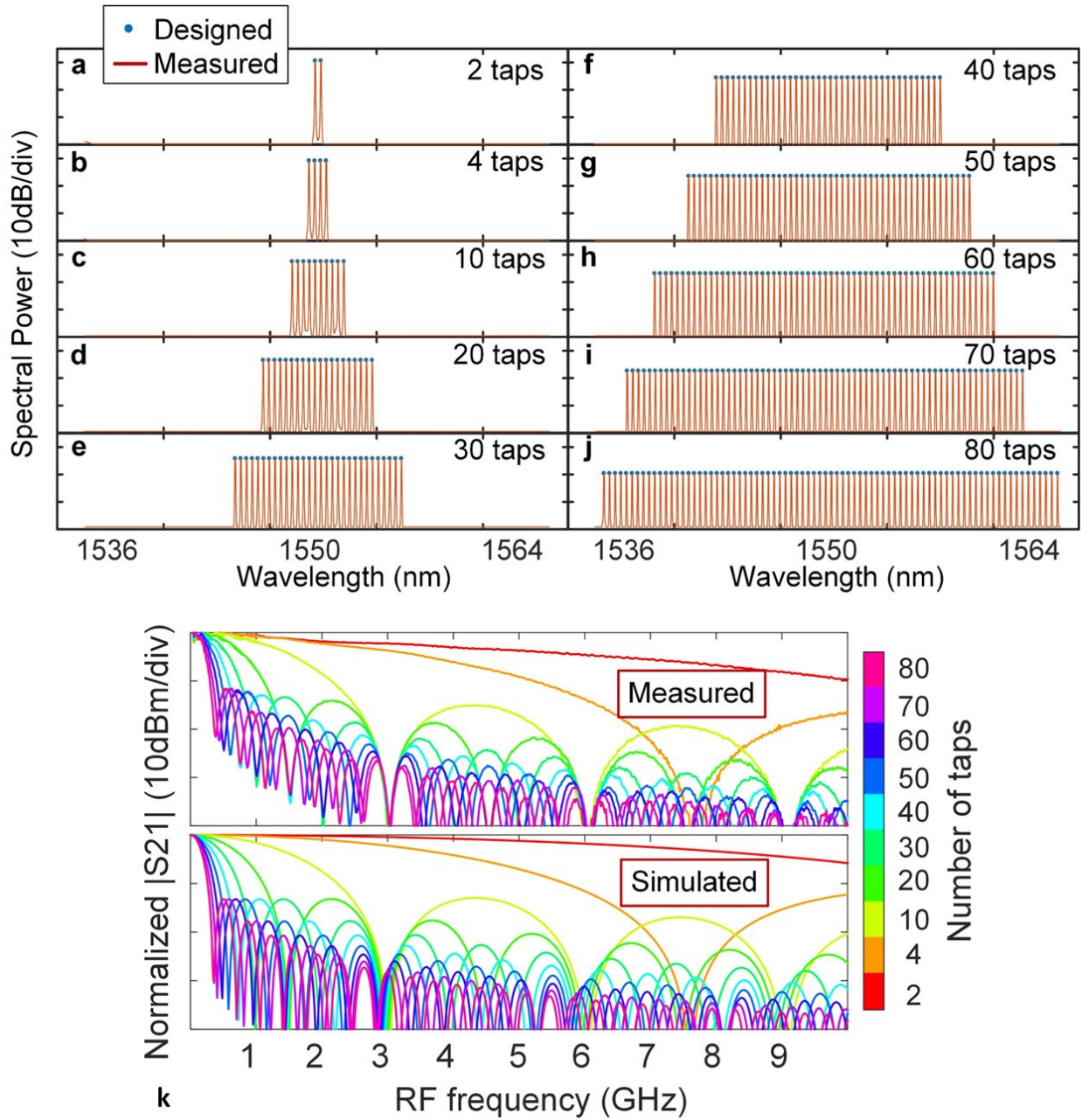

Fig. 6. Fixed RF lowpass filters using the 49GHz FSR microcomb with up to 80 uniform tap weights. (a-j) Optical spectra of the shaped micro-comb corresponding to the tap weights of a sinc filter with different tap number. (k) Measured and simulated RF transmission spectra of the sinc filter with different tap numbers.

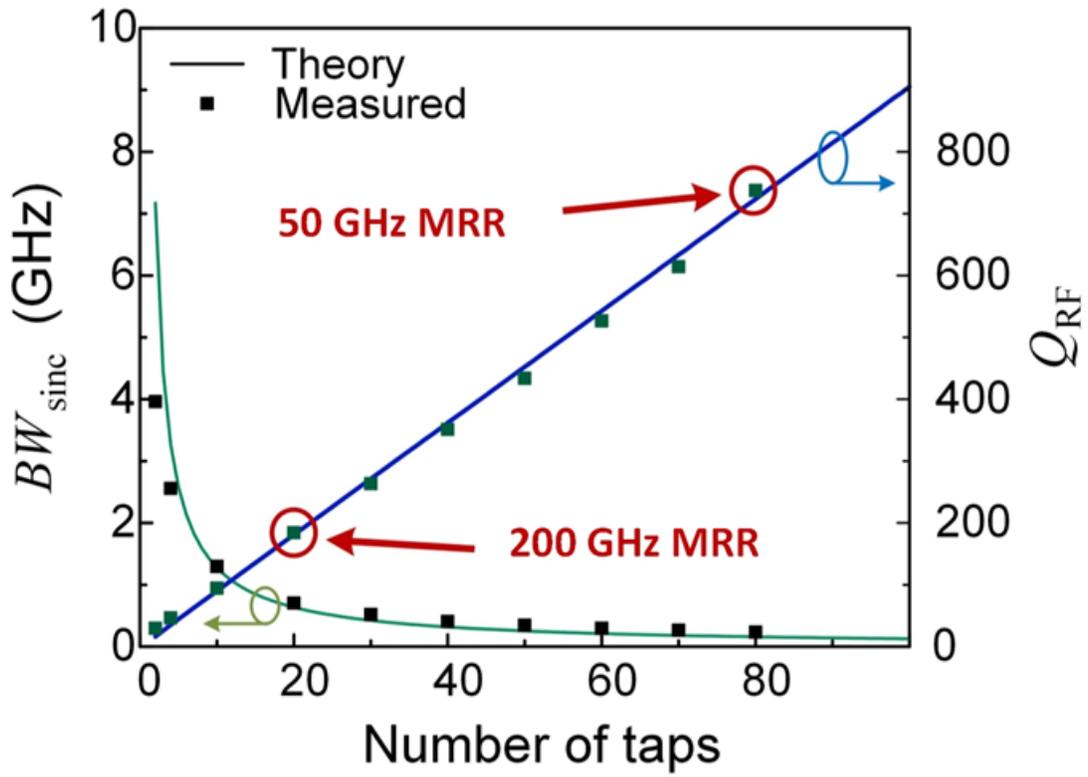

Fig. 7. Calculated (solid curves) and measured (dots) RF filter bandwidth and Q factor versus the number of taps, showing the results for the 200GHz FSR device and the 49GHz FSR comb, with the latter having up to 80 taps.

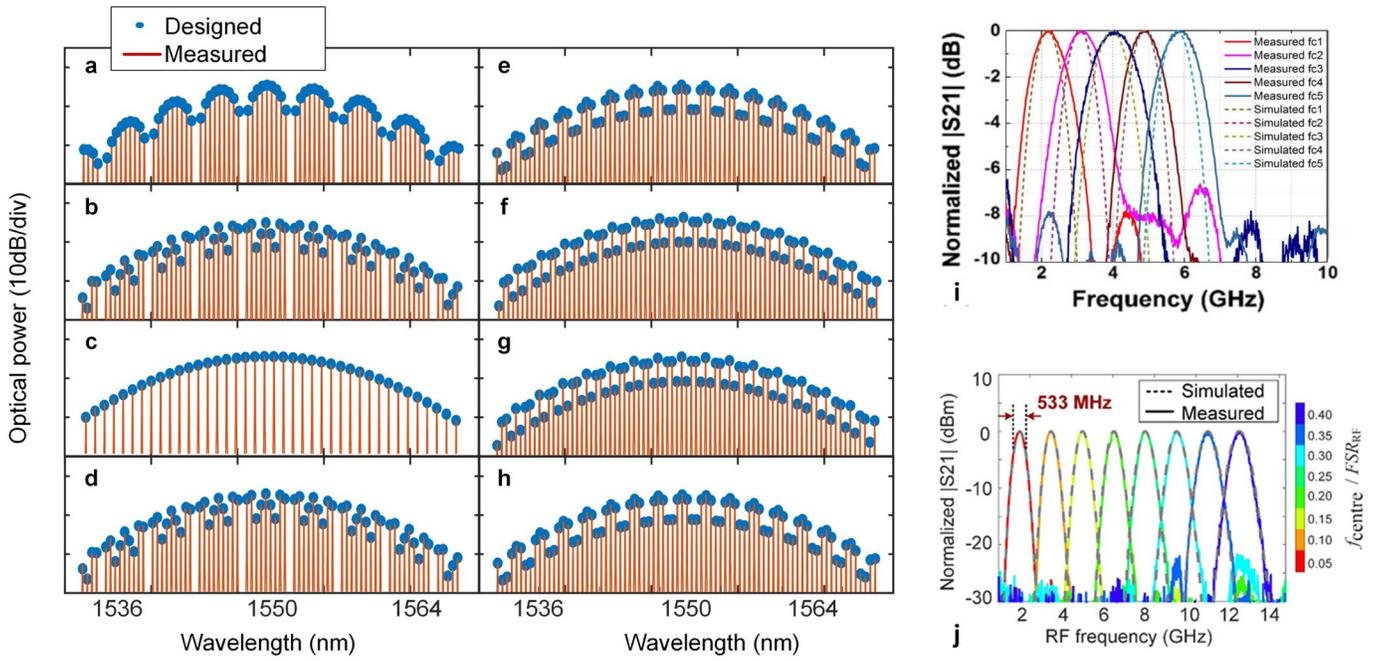

Fig. 8. (a-h) Shaped micro-comb corresponding to the centre-frequency-tunable sinc filter. RF transmission spectra of the for (i) 200GHz FSR device and (j) 49 GHz device based tunable RF filters.

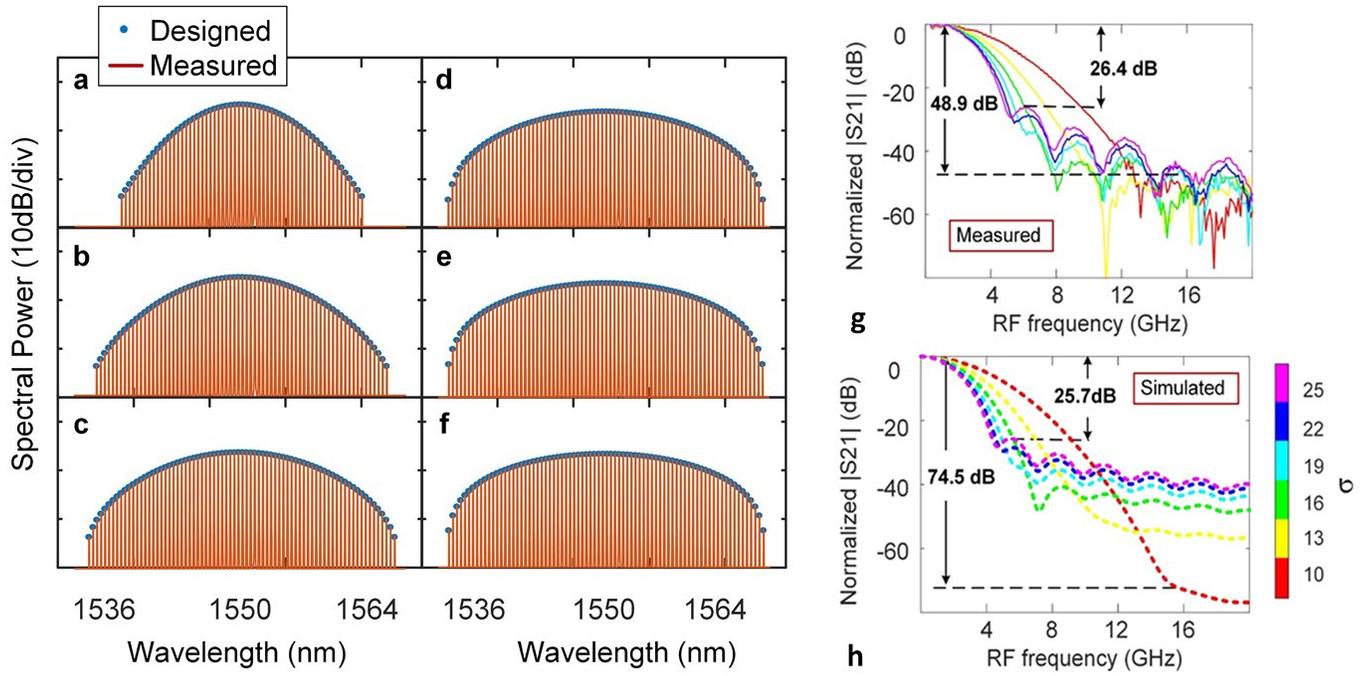

Fig. 9. Optical and RF transmission spectra of the apodized RF filter and corresponding taps weights using the 49GHz FSR comb. (a-f) The shaped comb spectra. (g) Measured and (h) simulated RF transmission spectra of the Gaussian-apodized sinc filter.

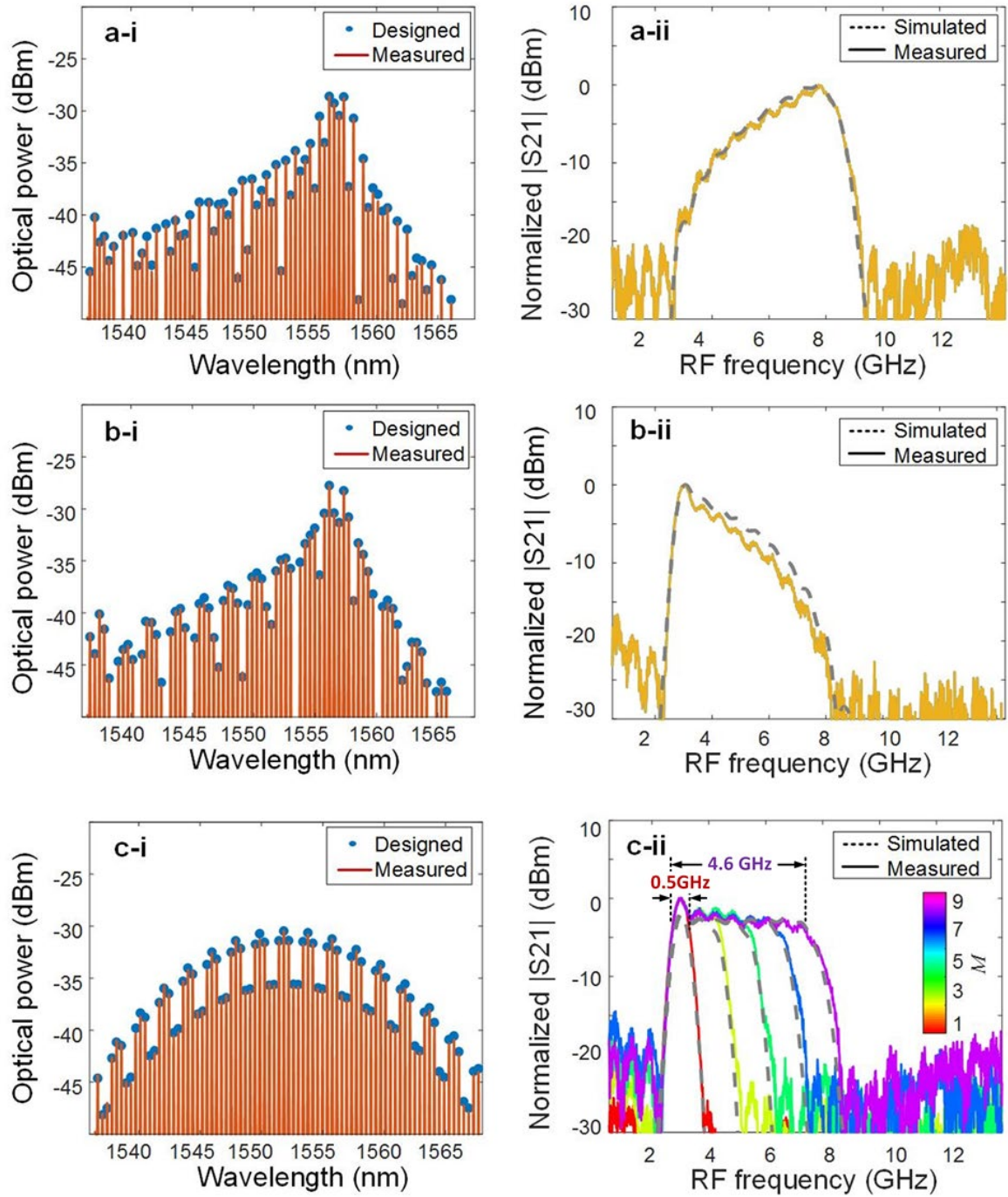

Fig. 10. Optical and RF transmission spectra for non-uniform gain equalizing filters, with (i) showing the tap weights and (ii) showing the RF transmission. All are based on the 49GHz FSR comb with 80 tap weights.